\documentstyle[12pt]{article}

\newcommand{\EQ}{\begin{equation}}
\newcommand{\EN}{\end{equation}}
\newcommand{\bea}{\begin{eqnarray}}
\newcommand{\ena}{\end{eqnarray}}
\newcommand{\vs}[1]{\vspace{#1 mm}}
\newcommand{\hs}[1]{\hspace{#1 mm}}
\renewcommand{\a}{\alpha}
\renewcommand{\b}{\beta}
\renewcommand{\c}{\gamma}
\renewcommand{\d}{\delta}
\newcommand{\e}{\epsilon}
\newcommand{\C}{\Gamma}
\def\bbox{{\,\lower0.9pt\vbox{\hrule \hbox{\vrule height 0.2 cm
\hskip 0.2 cm \vrule height 0.2 cm}\hrule}\,}}
\newcommand{\dsl}{\pa \kern-0.5em /}

\newcommand{\vp}{\varphi}

\newcommand{\pa}{\partial}
\renewcommand{\t}{\theta}
\newcommand{\tb}{{\bar \theta}}
\newcommand{\nn}{\nonumber\\}
\newcommand{\p}[1]{(\ref{#1})}

\begin{document}

\topmargin 0pt
\oddsidemargin 0mm
\renewcommand{\thefootnote}{\fnsymbol{footnote}}
\begin{titlepage}

\setcounter{page}{0}
\begin{flushright}
OU-HET 313 \\
hep-th/9902122
\end{flushright}

\vs{10}
\begin{center}
{\Large\bf Spacetime Superalgebra in AdS$_4 \times$ S$^7$
via Supermembrane Probe}
\vs{15}

{\large
Kazuyuki Furuuchi,\footnote{e-mail address:
furu@het.phys.sci.osaka-u.ac.jp}
Nobuyoshi Ohta\footnote{e-mail address:
ohta@phys.sci.osaka-u.ac.jp}
and
Jian-Ge Zhou\footnote{e-mail address: jgzhou@het.phys.sci.osaka-u.ac.jp,
JSPS postdoctoral fellow}}

\vs{10}
{\em Department of Physics, Osaka University,
Toyonaka, Osaka 560-0043, Japan} \\
\end{center}

\vs{15}
\centerline{{\bf{Abstract}}}
\vs{5}

The spacetime superalgebra via the supermembrane probe in the background of
AdS$_4 \times S^7$ is discussed to the lowest order in the spinor coordinate
$\t$. To obtain the correct spacetime superalgebras, all $\t^2$ order
corrections for supervielbein and super 3-form gauge potential have to be
included. The central extension of the superalgebra $OSp(8|4)$
of the super isometries for AdS$_4 \times S^7$ is found.

\end{titlepage}
\newpage

\renewcommand{\thefootnote}{\arabic{footnote}}
\setcounter{footnote}{0}

The subject of branes, singletons and superconformal field theories (SCFT)
on the anti-de Sitter (AdS) boundary was an active area of research about a
decade ago. For a review, see~\cite{DUF}. Recently there has been a renewed
interest in the AdS space by the conjectured AdS/CFT duality~\cite{M}. It is
known that the worldvolume dynamics of super $p$-branes propagating in AdS
backgrounds gives rise to $(p+1)$-dimensional SCFT~\cite{DUF}.
Thus from the connection between the near-horizon geometry of the
brane solutions in string theories and certain SCFT, it is of much
interest to study the supermembrane in the background of AdS$_4 \times S^7$.
Among others, progress has been made
recently~\cite{2}-\cite{13} in understanding Green-Schwarz superstring,
the D3-brane in the IIB supergravity background of AdS$_5 \times S^5$,
and supermembrane in $11D$ supergravity backgrounds of AdS$_4 \times S^7$
and AdS$_7 \times S^4$. In $11D$ supergravity, the AdS$_4 \times S^7$ and
AdS$_7 \times S^4$ backgrounds leave 32 supersymmetries intact.
These backgrounds are related to the near-horizon geometries
corresponding to M2- and M5-brane configurations and thus are related
to possible SCFT in 3 and 6 dimensions with 16 supersymmetries.

On the other hand, the analysis of supersymmetry algebras is one of
the most powerful approaches to M-theory. There are two kinds of
supersymmetries: one is the spacetime superalgebra and the other is
the worldvolume one. In refs.~\cite{14,15}, the superalgebras were
derived from ``test" M-brane actions in flat background, and recently
it was extended to the backgrounds of M2- and M5-branes~\cite{16}.
Due to the simplicity of the Killing spinors in the backgrounds of
M2- and M5-branes, the author of ref.~\cite{16} was allowed to set half of
the Killing spinor $\e^-$ and the fermionic coordinate $\t^-$ to zero
from the beginning. This effectively makes the transverse coordinates
inert under the spacetime diffeomorphism transformation and considerably
simplifies the necessary calculations.

In the present paper, we discuss how to construct the spacetime
superalgebras via supermembrane probe in the background of AdS$_4 \times S^7$.
In this geometry, there remain 32 intact supersymmetries in AdS$_4 \times S^7$
and the corresponding Killing spinors take the special form of a
triangular matrix with the vanishing lower left corner~\cite{6,17}.
So we cannot put $\e^-$ and $\t^-$ to zero, and should use the explicit
Killing spinors in AdS$_4 \times S^7$ to construct the
supercharges which possess the fermionic super isometry of the
background~\cite{12}. To get correct spacetime superalgebras
via supermembrane in AdS$_4 \times S^7$ in the Wess-Zumino-type
gauge~\cite{18}, we find that all $\t^2$ order corrections
for supervielbein and super 3-form gauge potential should
be included, which is different from the case in the background
of M2-branes~\cite{16} where some $\t^2$ order terms were ignored.
The reason is that the transverse coordinates inert under the superspace
diffeomorphism transformation are now activated in AdS$_4 \times S^7$
due to the Killing spinors characterizing the fermionic super isometry of
the background. Though this makes the whole calculations quite
complicated, we derive the explicit expressions for the supercharges.
Since there is ordering ambiguity in the definition of the
supercharges at the quantum level, we analyze the spacetime
superalgebras by computing the Poisson brackets of supercharges $\hat{Q}_\a$.
We find that the superalgebra $OSp(8|4)$ of super isometries of AdS$_4 \times
S^7$ is centrally extended. In the static membrane configurations (to be
defined later), the obtained Poisson brackets
$\{ \hat{Q}_\a , \hat{Q}_\b \}_{P.B.}$ can be
simplified, from which we can see that half of the supersymmetry is broken.
In other words, the superconformal symmetry generated by its generator
$S_\a$ is nonlinearly realized on the worldvolume fields. This means
that the superconformal symmetry is spontaneously broken, but it can be
recovered when the supermembrane sits on the boundary.

Now let us consider the near-horizon geometry of the M2-brane solution,
which is given by AdS$_4 \times S^7$:
\bea
ds^2 = \frac{R^2}{4}(dr^2 + e^{2r}\eta_{\mu \nu}dx^{\mu}dx^{\nu})
 + R^2 d\Omega^2_7, \label{m2} \nn
C^{(3)} = \left( \frac{R e^r}{2} \right)^3 dx^0\wedge dx^1\wedge dx^2,
\label{bg}
\ena
where $R \equiv l_P(2^5\pi^2N)^{1/6}$, $N$ is the number of M2-branes
and we have rescaled the coordinates of AdS$_4$ to make them dimensionless.
The Killing spinors satisfy the equations
\EQ
\d \psi_m = D_m \e + T{_m}^{n_1n_2n_3n_4} F_{n_1n_2n_3n_4} \e = 0,
\label{kil}
\EN
where $D_m = \pa_m + \frac{1}{4} \omega_m^{\hat{s}\hat{t}}\C_{\hat{s}\hat{t}}$
is the covariant derivative with respect to the local Lorentz transformation
and $T{_m}^{n_1n_2n_3n_4} = - \frac{1}{288} \left(\C{_m}^{n_1n_2n_3n_4} +
8 \C^{[n_1n_2n_3} \d^{n_4]}_m \right)$. In the AdS$_4 \times S^7$
background, the solutions to eq.~\p{kil} can be written as~\cite{17,9}
\EQ
\e = e^{-\frac{r}{2}\C} e^{\frac{\vp_{\hat{9}} }{2} \C \C_{\hat{r}}
 \C_{\hat{9}}}
\left( \prod_{\hat{a} = \hat{3}}^{\hat{8}}
 e^{-\frac{\vp_{\hat{a}}}{2}\C_{\hat{a},\hat{a}+\hat{1}}} \right)
\left[ 1-\frac{1}{2}x^{\mu}\d_{\mu}^{\hat{\mu}}
 \C_{\hat{\mu}}\C_{\hat{r}} \left( 1-\C \right) \right] \e_0 ,
\EN
where the factors with larger $\hat{a}$ values in
$\left( \prod_{\hat{a} = \hat{3}}^{\hat{8}}
 e^{-\frac{\vp_{\hat{a}}}{2}\C_{\hat{a},\hat{a}+\hat{1}}} \right)$
sit to the left of those with smaller $\hat{a}$ values,
$\C \equiv \C_{\hat{0}\hat{1}\hat{2}}$, $\e_0$ is an arbitrary
32-component constant Majorana spinor in $D=11$, and $\mu = 0,1,2$.
The coordinates $\vp^{\hat{a}} (\hat{a}=\hat{3}, \cdots,\hat{9})$
on the 7-sphere are defined iteratively by $d\Omega^2_{\hat a}=
d\vp_{\hat a}^2 + \sin^2 \vp_{\hat a}d\Omega_{\hat{a}-\hat{1}}^2$ with
$d\Omega^2_{\hat 3} = d\vp_{\hat 3}^2$.\footnote{We use the conventions in
ref.~\cite{16}:
$M,N$ denote superspace indices, $m,n$ curved ones, $\a,\b$ spinor ones,
$\mu,\nu$ coordinates parallel to the branes, $a,b$ coordinates transverse
to the branes, and tangent space indices are represented with hats on them.
Gamma matrices satisfy $\{ \C_{\hat m}, \C_{\hat n} \} = \eta_{\hat{m}\hat{n}}$
with $\eta_{\hat{m}\hat{n}} = {\rm diag.} (-,+,\cdots,+)$, and the Dirac
conjugate is defined by ${\bar \psi} = \psi^\dagger \C_0$. The worldvolume
chirality is denoted as $\psi^\pm \equiv \frac{1 \pm \C}{2} \psi$.}

The supermembrane action in $D=11$ supergravity background is~\cite{19}
\EQ
S = \int d^3\xi \left[ -\sqrt{-g\left( Z(\xi)\right)}
 + \frac{1}{3!} \e^{ijk}C^{(3)}_{ijk}  \right],
\label{sm}
\EN
where the superspace embedding coordinates are defined by
$Z^M(\xi) = \left( x^m({\xi}), \t^{\a}(\xi) \right)$ which are functions of
the worldvolume coordinates $\xi^i (i=0,1,2)$ and $\e^{012} = - \e_{012} = 1$.
The induced metric is defined by
$g_{ij} = \pa_i x^m(\xi) \pa_j x^n(\xi) \eta_{mn}$.

By gauge completion procedure~\cite{18} (or by the coset space construction
of AdS$_4 \times S^7$~\cite{8}), the supervielbein $E_M^{\hat{N}}$
in the Wess-Zumino-type gauge can be expressed up to terms of order $\t^2$ as
\bea
E_m^{\hat{n}} &=& e_m^{\hat{n}} - i\tb\C^{\hat{n}}
\left[ \frac{1}{4}\omega_m^{\hat{s}\hat{t}}\C_{\hat{s}\hat{t}}
+ T{_m}^{n_1n_2n_3n_4}F_{n_1n_2n_3n_4} \right]
\t + O(\t^3), \nn
E_m^{\hat{\a}} &=& \frac{1}{4}\omega_m^{\hat{s}\hat{t}}
(\C_{\hat{s}\hat{t}}\t)^{\hat{\a}}
+ (T{_m}^{n_1n_2n_3n_4}\t)^{\hat{\a}}F_{n_1n_2n_3n_4} + O(\t^2), \nn
E_\a^{\hat{m}} &=& i(\tb\C^{\hat{m}})_\a + O(\t^3), \nn
E_\a^{\hat{\b}} &=& \d_\a^{\hat{\b}} + M_\a^{\hat{\b}} + O(\t^3),
\label{sv}
\ena
where $M_\a^{\hat{\b}}$ represent the $F\t^2$ terms, which do not
affect our following discussions. In eq.~\p{sv}, the fermionic terms are
suppressed for our background~\p{bg}. In ref.~\cite{16}, the terms at order
$\t^2$ in $E_M^{\hat{N}}$ were ignored because the transverse coordinates
are inert under superspace diffeomorphism transformation. In the superspace,
the super-coordinate transformation can be defined as~\cite{20}
\bea
&& \d Z^M = \Xi^M, \nn
&& \Xi^{\mu} = i\bar{\e}\C^{\mu}\t , \; \;
\Xi^a = i\bar{\e}\C^a\t , \; \;
\Xi^r = i\bar{\e}\C^r \t , \; \;
\Xi^{\a} = \e^\a .
\label{sc}
\ena

The Wess-Zumino Lagrangian ${\cal L}_{WZ}$ is given by~\cite{18}
\bea
{\cal L}_{WZ} &=& \frac{1}{3!}\e^{ijk}C^{(3)}_{ijk}  \nn
&=& -\frac{1}{6}\e^{ijk}\pa_ix^{m_1}\pa_jx^{m_2}\pa_jx^{m_3} \nn
&& \times \left[
C^{(3)}_{m_1m_2m_3}
+ \frac{3i}{4}\tb\C_{\hat{n}_1\hat{n}_2}
\C_{m_1m_2}\t\omega_{m_3}^{\hat{n}_1\hat{n}_2}
+ 3i\tb\C_{m_1m_2}T{_m}^{n_1n_2n_3n_4}\t F_{n_1n_2n_3n_4}
\right] \nn
&&-\frac{i}{2}\e^{ijk}\pa_ix^{m_1}\pa_jx^{m_2}
\tb\C_{m_1m_2}\pa_k\t + O(\t^3).
\label{wz}
\ena

Under the infinitesimal superspace transformation \p{sc}, the variation
of ${\cal L}_{WZ}$ is found to be
\bea
\d {\cal L}_{WZ} &=& - \frac{i}{2}\e^{ijk}\pa_i
 \left\{ \pa_jx^{\mu}\pa_kx^{\nu} \bar{\e}
 \left[ \left( \frac{R e^r}{2} \right)^3 \e_{\mu\nu\rho}\C^{\rho}
+ \C_{\mu\nu} \right] \t
+ 2\pa_jx^{\mu}\pa_k r \bar{\e}\C_{\mu r}\t \right.  \nn
&&  \; \; \; \; \;
 + 2\pa_jx^{\mu}\pa_k\vp^a \bar{\e}\C_{\mu a}\t
 + 2\pa_j r \pa_k\vp^a \bar{\e}\C_{ra}\t
 + \pa_j\vp^a\pa_k\vp^b \bar{\e}\C_{ab}\t \Biggr\} + O(\t^3),
\label{dwz}
\ena
which means that the supermembrane action \p{sm} is invariant under the
super-coordinate transformation \p{sc} up to total derivatives.
In deriving eq.~\p{dwz}, we have used eqs.~\p{m2} and \p{sv}-\p{wz}.

Now we use the explicit Killing spinors in AdS$_4 \times S^7$ to
construct the supercharges, which possess the fermionic super isometry
of the background~\cite{12}. In the Hamiltonian formalism, the Noether
supercharges can be defined as an integral over the test membrane at
a fixed time~\cite{14}. We get
\bea
\hat{Q}_\a &=& \int d^2\xi \left\{ -\left( \left[
 1+\frac{1}{2}x^{\nu}\d_{\nu}^{\hat{\nu}} \C_{\hat{\nu}}\C_{\hat{r}}
 \left( 1+\C \right) \right] \biggl(
   \prod_{\hat{a} = \hat{3}}^{\hat{8}}
    e^{\frac{\vp_{\hat{a}}}{2}\C_{\hat{a},\hat{a}+\hat{1}}} \biggr)
\right. \right. \nn
&& \; \;\times 
 e^{-\frac{\vp_{\hat{9}}}{2}\C\C_{\hat{r}}\C_{\hat{9}}}
e^{-\frac{r}{2}\C}
\left( \C^{\mu}\Pi_{\mu} + \C^{r}\Pi_{r} + \C^a\Pi_a \right)
\Biggr)_{\a\b}\t^{\b} \nn
&& + i \left(
 e^{-\frac{r}{2}\C} e^{\frac{\vp_{\hat{9}}}{2}\C\C_{\hat{r}}\C_{\hat{9}}}
 \biggl( \prod_{\hat{a} = \hat{3}}^{\hat{8}}
 e^{-\frac{\vp_{\hat{a}}}{2}\C_{\hat{a},\hat{a}+\hat{1}}}  \biggr) 
\left[ 1-\frac{1}{2}x^{\nu}\d_{\nu}^{\hat{\nu}} \C_{\hat{\nu}}\C_{\hat{r}}
 \left( 1-\C \right) \right]\C_{\hat{0}}
 \right){^\b}_{\a}\Pi_\b  \nn
&& - \frac{1}{2}\e^{0ij} \left(
 \left[ 1+\frac{1}{2}x^{\sigma}\d_{\sigma}^{\hat{\sigma}}
 \C_{\hat{\sigma}}\C_{\hat{r}} \left( 1+\C \right) \right]
\biggl( \prod_{\hat{a} = \hat{3}}^{\hat{8}}
 e^{\frac{\vp_{\hat{a}}}{2}\C_{\hat{a},\hat{a}+\hat{1}}} \biggr)
 e^{-\frac{\vp_{\hat{9}}}{2}\C\C_{\hat{r}}\C_{\hat{9}}}
 e^{-\frac{r}{2}\C} \right. \nn
&& \; \; \times  \left[
 \pa_ix^{\mu}\pa_jx^{\nu} \biggl(\frac{R^3}{8}e^{3r}
 \e_{\mu\nu\rho}\C^{\rho} + \C_{\mu\nu} \biggr)
 + 2\pa_ix^{\mu}\pa_j r \C_{\mu r} \right.  \nn
&& \left. \left. \; \; \; \; \;
 + 2\pa_ix^{\mu}\pa_j\vp^b \C_{\mu b}
 + 2\pa_i r \pa_j\vp^b \C_{rb} + \pa_i\vp^b\pa_j\vp^c \C_{bc} \Biggr]
  \right)_{\a\b}\t^\b \right\} + O(\t^3),
\label{qa}
\ena
where $\Pi_\mu, \Pi_\c, \Pi_a$ and $\Pi_\a$ are conjugate momenta of
$x^{\mu}, r, \vp^a$ and $\t^\a$, respectively. In deriving eq.~\p{qa},
we have included all the $\t^2$ order corrections for supervielbein
and super 3-form gauge potential, which is different from the approximation
in ref.~\cite{16} where some $\t^2$ order terms were ignored.

Up to this point, we have obtained the supercharges via supermembrane
probe in AdS$_4 \times S^7$, which looks quite complicated. However,
due to the special structure of the supercharges, we can simplify the
expression in certain gauge. Since there is ordering ambiguity in the
definition of supercharges \p{qa} at the quantum level, we restrict
the following discussions to the classical level. The generalization
to the quantum level is left to future work.

The spacetime superalgebras of $\hat{Q}_\a$ can be obtained as
\bea
\left\{ \hat{Q}_\a , \hat{Q}_\b \right\}_{P.B.} \hs{-5}
&&= -i \int d^2\xi \Biggl\{ \Pi_\mu \left(
 2e^{-r \C}\C^{\mu}\C_{\hat{0}} - 2e^{-r}x^{\nu}\d_{\nu}^{\hat{\nu}}
 \C{^\mu}_{\hat{\nu}} \C \C_{\hat{r}}\C_{\hat{0}}
 - 2e^{-r}x^{\mu}\C_{\hat r}\C_{\hat{0}} \right. \nn
&& \left. + e^{-r} \left[ x^2\C^{\mu}
 - 2x^{\mu}(x^{\nu}\d_{\nu}^{\hat{\nu}} \C_{\hat \nu}) \right]
 \left(1-\C \right) \C_{\hat{0}} \right)
+ 4\Pi_r \left[ \C_{\hat{r}} + x^{\nu}\d_{\nu}^{\hat{\nu}} \C_{\hat{\nu}}
  \left( 1-\C \right) \right] \C_{\hat{0}} R^{-1} \nn
&& \left. + 2\Pi_a \biggl(
 \prod_{\hat{b} = \hat{3}}^{\hat{8}}
 e^{\frac{\vp_{\hat{b}}}{2}\C_{\hat{b},\hat{b}+\hat{1}}} \biggr)
 e^{-\frac{\vp_{\hat{9}}}{2}\C\C_{\hat{r}}\C_{\hat{9}}} \C^a\C_{\hat{0}}
 e^{ \frac{\vp_{\hat{9}}}{2}\C\C_{\hat{r}}\C_{\hat{9}}}
 \biggl( \prod_{\hat{c} = \hat{3}}^{\hat{8}}
 e^{-\frac{\vp_{\hat{c}}}{2}\C_{\hat{c},\hat{c}+\hat{1}}} \biggr)
 \right\}_{\a\b} \nn
&& + C^{(M2)}_{\a\b} + O(\t^2) ,
\label{qalg}
\ena
where the central charge $C^{(M2)}_{\a\b}$ originating from the total
derivative terms in the variation of supermembrane action under the
infinitesimal superspace diffeomorphism transformation \p{sc} is
given by
\EQ
C^{(M2)} = B_{WZ} + B^T_{WZ},
\EN
with
\bea
B_{WZ} &=& \int \d^2\xi \; \frac{i}{2} \e^{0ij}
 \left\{ \left[ 1+\frac{1}{2} x^{\rho}\d_{\rho}^{\hat{\rho}}
 \C_{\hat{\rho}} \C_{\hat{r}} \left( 1 + \C \right) \right]
 \left( \prod_{\hat{a} = \hat{3}}^{\hat{8}}
 e^{\frac{\vp_{\hat{a}}}{2}\C_{\hat{a},\hat{a}+\hat{1}}} \right)
 e^{-\frac{\vp_{\hat{9}}}{2}\C\C_{\hat{r}}\C_{\hat{9}}} e^{-\frac{r}{2}\C}
\right. \nn
&&
\times \biggl[ \pa_ix^{\mu}\pa_jx^{\nu} \biggl(
 \frac{R^3}{8}e^{3r}\e_{\mu\nu\rho}\C^{\rho} + \C_{\mu\nu} \biggr) \nn
&& + 2\pa_ix^{\mu} \pa_j r \C_{\mu r} + 2\pa_ix^{\mu} \pa_j\vp^b
 \C_{\mu b} + 2\pa_i r \pa_j\vp^b \C_{rb}
 + \pa_i\vp^b \pa_j\vp^c \C_{bc} \biggr] \nn
&& \left. \times
 e^{-\frac{r}{2}\C} e^{\frac{\vp_{\hat{9}}}{2}\C\C_{\hat{r}}\C_{\hat{9}}}
\left( \prod_{\hat{a} = \hat{3}}^{\hat{8}}
 e^{-\frac{\vp_{\hat{a}}}{2}\C_{\hat{a},\hat{a}+\hat{1}}} \right)
\left[ 1 - \frac{1}{2}x^{\sigma}\d_{\sigma}^{\hat{\sigma}}
 \C_{\hat{\sigma}}\C_{\hat r} \left( 1-\C \right) \right] \C_{\hat{0}} \right\}.
\label{bwz}
\ena
This is the general result we obtain. We will see that it can be
extremely simplified in certain gauge.

If we decompose the 11-dimensional gamma matrices into the ones
referring to the AdS$_4$ and $S^7$ spaces and denote them by $\c^s$
and ${\c'}_{s'}$ respectively, eq.~\p{qalg} can be written as
\bea
\left\{ \hat{Q}_{\a\a'} , \hat{Q}_{\b\b'} \right\}_{P.B.}
&=& - \left( \c_5 C \right)_{\a\b}
 \left[ 2 \left( {\c '}_{s'} C'  \right)_{\a'\b'} P^{s'}
 - h \left(  {\c'}_{s't'} C' \right)_{\a'\b'} M^{s't'} \right] \nn
&& - {C'}_{\a'\b'}
 \left[ 2 \left( \c_s C  \right)_{\a\b} P^{s}
 + 2h \left(  \c_{st} \c_5 C \right)_{\a\b} M^{st} \right] \nn
&& + C^{(M2)}_{\a\b,\a'\b'}
 + O(\t^2),
\label{qq}
\ena
where $\c_5 = \c_0\c_1\c_2\c_r$ and the charge conjugation matrix
${\cal C}$ is decomposed as ${\cal C} = C \otimes C'$. The coefficient
$h$ is read off from $F_{\mu\nu\rho\sigma} = 6he \e_{\mu\nu\rho\sigma}$
which in turn can be obtained from eq.~\p{bg}. The $P_s$ and $M_{st}$
correspond to the bosonic generators in $SO(3,2)$ while $P_{s'}$ and
$M_{s't'}$ to those in $SO(8)$~\cite{2}. The Poisson bracket~\p{qq} of
the supercharges $\hat{Q}_\a$, which are induced from the worldvolume of
supermembrane probe in AdS$_4 \times S^7$, shows that the superalgebra
$OSp(8|4)$ of the super isometries of the background is centrally extended.
The complete central extension of the superalgebra $OSp(8|4)$ by
explicit field construction in the worldvolume of the supermembrane
will be discussed elsewhere.

Now let us see how the above superalgebras of supercharges $\hat{Q}_\a$ can
be simplified in the static membrane configuration, from which we can read
off the unbroken supersymmetries. It is defined by~\cite{11}
\EQ
x^{\mu} = \xi^{\mu}, \; \;
r = {\rm const.} \; ,\; \;
\vp^a = {\rm const.} \; ,\; \;
\t^{\a} = 0,
\label{sta}
\EN
for which we have
\bea
\Pi_\mu = \frac{\d L^{(0)}}{\d\dot{x}^{\mu}}
 + \frac{1}{2}\e^{0ij} \pa_i x^{\nu}\pa_j x^{\rho}
 C^{(3)}_{\mu\nu\rho} = 0, \; \;
\Pi_{\c} = \Pi_a = 0,
\label{pi}
\ena
where $L^{(0)}$ is the first term in eq.~\p{sm}.
With the help of eqs.~\p{sta} and \p{pi}, we can simplify eq.~\p{qq} to obtain
\EQ
\left\{ \hat{Q}_\a , \hat{Q}_\b \right\}_{P.B.}
= i \int d^2\xi \; \frac{R^2}{2} \left[ e^{(2-\C)r}
\left( 1 - \C \right) \right]_{\a\b},
\EN
from which we see that in the static membrane configuration~\p{sta}, half of
the supersymmetry is broken. In fact, if we choose diagonal representation for
the matrix $\C = \mbox{diag} ({\bf 1}_{16 \times 16},-{\bf 1}_{16 \times 16})$
and decompose $\hat{Q}_\a$ into $Q_\a = \hat{Q}_{+\a} \equiv \frac{1+\C}{2}
\hat{Q}_\a$, $S_\a = \hat{Q}_{-\a} \equiv  \frac{1-\C}{2} \hat{Q}_\a$, which
correspond to Poincar\'{e} and conformal supersymmetry generators~\cite{11},
we have
\bea
\lefteqn{ \{ Q_\a , Q_\b \}_{P.B.} = 0 , \; \; \;
\{ Q_\a , S_\a \}_{P.B.} = 0 , } \nn
&&\{ S_\a , S_\b \}_{P.B.}
 = i \int d^2\xi \frac{R^2}{2} e^{3r} \d_{\a\b}.
\label{qs}
\ena
Eq.~\p{qs} shows that the superconformal symmetry generated by $S_\a$ is
nonlinearly realized on the worldvolume fields, which means that if the
supermembrane probe sits at any finite $r$, the manifest superconformal
symmetry is broken when we gauge-fix the action and express it in terms
of the physical transverse coordinate fields. However, the superconformal
symmetry is recovered when the supermembrane sits on the boundary at
$r \to -\infty$, which is consistent with the known results~\cite{2,11,12}.

In the above discussion, we have only constructed the supercharges in the
Wess-Zumino-type gauge to the lowest order in $\t$. It would be interesting
to check whether it is possible to construct them in Killing spinor
gauge~\cite{4,6} to all orders in $\t$. It is also interesting to discuss
the spacetime superalgebras via super 5-brane probes. One of the interesting
examples is to consider the construction via D5-brane probe in
AdS$_5\times S^5$, and to see how to realize the maximally extended D5-brane
worldvolume supersymmetry algebra from the explicit superalgebras~\cite{21}.

\section*{Acknowledgement}

We would like to thank M.J. Duff for helpful comments.
This work was supported in part by the grant-in-aid from the Ministry of
Education, Science, Sports and Culture No. 96208.

\newcommand{\NP}[1]{Nucl.\ Phys.\ {\bf #1}}
\newcommand{\AP}[1]{Ann.\ Phys.\ {\bf #1}}
\newcommand{\PL}[1]{Phys.\ Lett.\ {\bf #1}}
\newcommand{\CQG}[1]{Class. Quant. Gravity {\bf #1}}
\newcommand{\CMP}[1]{Comm.\ Math.\ Phys.\ {\bf #1}}
\newcommand{\PR}[1]{Phys.\ Rev.\ {\bf #1}}
\newcommand{\PRL}[1]{Phys.\ Rev.\ Lett.\ {\bf #1}}
\newcommand{\PRE}[1]{Phys.\ Rep.\ {\bf #1}}
\newcommand{\PTP}[1]{Prog.\ Theor.\ Phys.\ {\bf #1}}
\newcommand{\PTPS}[1]{Prog.\ Theor.\ Phys.\ Suppl.\ {\bf #1}}
\newcommand{\MPL}[1]{Mod.\ Phys.\ Lett.\ {\bf #1}}
\newcommand{\IJMP}[1]{Int.\ Jour.\ Mod.\ Phys.\ {\bf #1}}
\newcommand{\JHEP}[1]{J.\ High\ Energy\ Phys.\ {\bf #1}}
\newcommand{\JP}[1]{Jour.\ Phys.\ {\bf #1}}

\end{document}